%% file: main.tex
\documentclass[conference]{IEEEtran}
\IEEEoverridecommandlockouts
\usepackage{cite}
\usepackage{amsmath,amssymb,amsfonts}
\usepackage{algorithmic}
\usepackage{graphicx}
\usepackage{textcomp}
\usepackage{xcolor}
\def\BibTeX{{\rm B\kern-.05em{\sc i\kern-.025em b}\kern-.08em
    T\kern-.1667em\lower.7ex\hbox{E}\kern-.125emX}}

\usepackage{booktabs}
\usepackage{longtable}
\usepackage{threeparttable}

\usepackage{xspace}
\usepackage{xcolor}
\usepackage{soul}
\usepackage{graphicx}
\usepackage{colortbl}
\usepackage[most]{tcolorbox}

\usepackage{listings}
\usepackage{verbatimbox}
\usepackage{booktabs}
\usepackage{multirow} 
\usepackage{algorithmic}
\usepackage{graphicx}
\usepackage{multirow}
\usepackage{color,soul}
\usepackage{textcomp}
\usepackage{xcolor}
\usepackage{todonotes}
\usepackage{listings}
\usepackage{wrapfig}
\usepackage{tabularx}
\usepackage{url}
\usepackage{caption}
\usepackage{subcaption}

\usepackage{beramono}

\input{macro}

\DeclareRobustCommand{\IEEEauthorrefmark}[1]{\smash{\textsuperscript{\footnotesize #1}}}
\begin{document}

\title{
Designing Conversational AI to Support Think-Aloud Practice in Technical Interview Preparation for CS Students
}



\author{
\IEEEauthorblockN{
Taufiq Daryanto\IEEEauthorrefmark{1},
Sophia Stil\IEEEauthorrefmark{1},
Xiaohan Ding\IEEEauthorrefmark{1},
Daniel Manesh\IEEEauthorrefmark{1},
Sang Won Lee\IEEEauthorrefmark{1},\\
Tim Lee\IEEEauthorrefmark{2},
Stephanie Lunn\IEEEauthorrefmark{3},
Sarah Rodriguez\IEEEauthorrefmark{1},
Chris Brown\IEEEauthorrefmark{1},
Eugenia H. Rho\IEEEauthorrefmark{1}
}
\IEEEauthorblockA{\IEEEauthorrefmark{1}Virginia Tech, Blacksburg, VA, USA\\
\{taufiqhd, ssophia, xiaohan, danielmanesh, sangwonlee, srodriguez, dcbrown, eugenia\}@vt.edu}
\IEEEauthorblockA{\IEEEauthorrefmark{2}CodePath, USA, tim@codepath.org}
\IEEEauthorblockA{\IEEEauthorrefmark{3}Florida International University, USA, sjlunn@fiu.edu}
}


\maketitle

\begin{abstract}
One challenge in technical interviews is the think-aloud process, where candidates verbalize their thought processes while solving coding tasks. Despite its importance, opportunities for structured practice remain limited. Conversational AI offers potential assistance, but limited research explores user perceptions of its role in think-aloud practice. To address this gap, we conducted a study with 17 participants using an LLM-based technical interview practice tool. Participants valued AI’s role in simulation, feedback, and learning from generated examples. Key design recommendations include promoting social presence in conversational AI for technical interview simulation, providing feedback beyond verbal content analysis, and enabling crowdsourced think-aloud examples through human-AI collaboration. Beyond feature design, we examined broader considerations, including intersectional challenges and potential strategies to address them, how AI-driven interview preparation could promote equitable learning in computing careers, and the need to rethink AI’s role in interview practice by suggesting a research direction that integrates human-AI collaboration.
\end{abstract}


\section{Introduction}
Employment in the field of computer and information technology is expected to grow at a rate significantly faster than the average for all occupations over the next decade \cite{bls2024}. During the hiring process for these computing roles, technical interviews remain a prevalent method for evaluating candidates \cite{ford2017tech, behroozi2019hiring}. A technical interview is a specific type of job interview that requires candidates to write code that solves particular programming challenges while simultaneously performing \textit{think-aloud} \cite{aziz2012elements}. Think-aloud, in this context, involves verbalizing one’s thinking process and problem-solving approach to the interviewers as they work through the coding task \cite{mcdowell2013cracking}. This practice allows interviewers to assess not only a candidate’s technical abilities but also their approach to problem-solving, clarity in communication, and adaptability under pressure \cite{ford2017tech}.

Despite its importance, the think-aloud process remains a significant source of anxiety for candidates due to its cognitive and performative demands \cite{behroozi2019hiring, behroozi2022asynchronous}, especially among computer science (CS) students \cite{behroozi2020does}. Preparation through mock interviews can alleviate these challenges, but such opportunities are not equally accessible for students \cite{lunn2022need}. Meanwhile, existing coding practice tools, such as LeetCode \cite{leetcode} and HackerRank \cite{hackerrank}, focus solely on solving coding problems without offering critical support for facilitating think-aloud practice. These gaps leave many candidates underprepared for real-world technical interviews \cite{ford2017tech}.

Conversational Artificial Intelligence (AI), particularly based on large language models (LLMs), has shown potential to bridge these gaps. Researchers have examined how LLM-powered conversational AI simulate general interview scenarios \cite{ajri2023virtual, daryanto2024conversate}, provide feedback on verbalized thought processes \cite{zhang2024using}, and offer insights into effective communication strategies \cite{shaikh2024rehearsal}. However, existing research has largely focused on general communication training \cite{dai2024generative, rusmiyanto2023role, stamer2023artificial, agrawal2024impact, liaw2023artificial, shorey2019virtual} and standard interview preparation \cite{daryanto2024conversate, hoque2013mach, TARDIS, inoue2020jobErica2, rasipuram2020automatic, naim2016automated, Chou2022-mm, thakkar2023automatic, li2023ezinterviewer}, leaving the domain of technical interview preparation underexplored. This gap is particularly critical as technical interviews remain a key requirement for computing jobs, demanding that candidates simultaneously solve coding problems and articulate their reasoning, a combination of skills that presents unique challenges not addressed in other domains \cite{mcdowell2013cracking}.

Our research aims to address this gap by investigating how LLM-based conversational AI can facilitate think-aloud practice for technical interview preparation. We focus on understanding user perceptions, identifying their concerns, and uncovering critical design considerations. Although some commercial AI tools have started to support technical interview preparation \cite{aptico}, specific design considerations required to effectively leverage LLMs for think-aloud practice remain largely unexplored. Gaining insights from users is essential to ensure these tools are not only effective but also aligned with their diverse needs and challenges \cite{kabir2022ask}. To address this, our paper inquires about user-centered insights to inform the development of conversational AI systems for think-aloud practice. Our study is guided by this main research question: \textbf{How do users perceive the role of conversational AI for think-aloud practice in technical interview preparation?}

To explore how people perceived AI's role in supporting think-aloud practice for technical interviews, we developed an LLM-based system for technical interview practice. Our tool had three core features: (1) technical interview simulation (Fig. \ref{fig:simulation_feature}), (2) AI feedback on think-aloud practice (Fig. \ref{fig:feedback_feature}), and (3) AI-generated think-aloud example dialogue (Fig. \ref{fig:example_feature}). We conducted a user study with 17 participants to explore their perceptions of AI-assisted think-aloud practice. Rather than evaluating the tool itself, our primary goal was to gather user insights, focusing on their preferences, concerns, and recommendations for leveraging AI in this context.

Our findings reveal several key insights into how conversational AI can transform technical interview preparation while surfacing critical tensions that must be addressed in its design. Participants perceived that AI simulations could create genuine technical interview experiences, valuing the two-way dialogue that helped them articulate complex technical reasoning in their think-aloud process. Additionally, our study also uncovered nuanced needs around feedback mechanisms. Participants emphasized that effective think-aloud preparation requires more than just analysis of verbal explanations. Participants sought AI guidance on balancing thinking, talking, and coding simultaneously. Moreover, while participants valued learning from AI-generated think-aloud example dialogues, they proposed an alternative approach by combining human authenticity with AI capabilities through human-AI collaborative crowdsourced examples.

Furthermore, beyond these feature-specific insights, we also discussed broader issues and considerations in developing conversational AI for technical interview preparation. For instance, we discussed the need to rethink the role of AI in technical interview practice, suggesting a research direction that leverages human-AI collaboration. Additionally, we examined a few intersectional challenges users faced and discussed potential strategies to address them. Overall, our main contributions from the study are: 1) A web-based LLM-based tool for supporting think-aloud practice in technical interview preparation; 2) Empirical insights into user perceptions of AI-assisted think-aloud practice; 3) Design considerations for developing AI-assisted technical interview preparation; 4) Broader implications for equitable access to technical interview preparation.






\section{Related Work}

\subsection{Challenges and Gaps in Think-Aloud Practice for Technical Interviews}

The think-aloud process is central to technical interviews, requiring candidates to verbalize their problem-solving approaches while simultaneously writing code \cite{behroozi2020does}. This method allows interviewers to assess not only technical proficiency but also reasoning, communication clarity, and adaptability under pressure \cite{mcdowell2013cracking, ford2017tech}. However, many candidates find the think-aloud component highly demanding due to its cognitive and performative requirements, which can induce significant anxiety \cite{behroozi2020does, behroozi2022asynchronous}.


Despite the importance of think-aloud training, practice opportunities to develop this skill remain limited. For instance, current computing courses in higher education rarely include opportunities for students to practice this critical skill \cite{lunn2022need, lunn2024educational}. Mock technical interviews are one proposed solution, as they enable candidates to practice verbalizing their thought processes while solving coding problems \cite{wilkie2024efficacy}. However, access to mock interviews is often limited, requiring skilled peers or mentors who are familiar with the technical interview process \cite{lunn2022need}. Commercial solutions, such as paid mock interview sessions with hiring managers, are prohibitively expensive, often costing upwards of \$150 per session \cite{igotAnOffer}, and rarely used by candidates~\cite{bell2025}.

Researchers have also examined the role of AI in programming education. For example, studies have explored how AI tools like ChatGPT support students in coding tasks such as debugging, answering conceptual questions, and generating complete coding solutions \cite{wang2023exploring, ghimire2024coding}. AI has also been leveraged to create educational resources \cite{guo2023six}, including automatically generated coding exercises to aid learning \cite{sarsa2022automatic, becker2023programming}. Additionally, researchers have developed conversational AI systems to personalize the learning process for specific computing skills \cite{jin2024teach}, such as debugging \cite{ma2024teach} and prompt engineering \cite{ma2024you}.

However, while those tools have shown promise in helping learners improve their technical skills by providing personalized and scalable solutions \cite{fenu2024exploring, wang2023exploring}, they often overlook the think-aloud component, which is an increasingly critical aspect of technical interviews \cite{ford2017tech}. Beyond coding skills, candidates are expected to articulate their thought processes and problem-solving strategies \cite{ford2017tech}, yet existing tools fail to address this dual need for coding and communication training \cite{leetcode, hackerrank, jin2024teach}. This gap warrants the need to explore systems that integrate coding practice with think-aloud training to align with industry requirements for assessing candidates through technical interviews \cite{ford2017tech, mcdowell2013cracking}.

\begin{figure*}[t] 
    \centering
    \includegraphics[width=.85\textwidth]{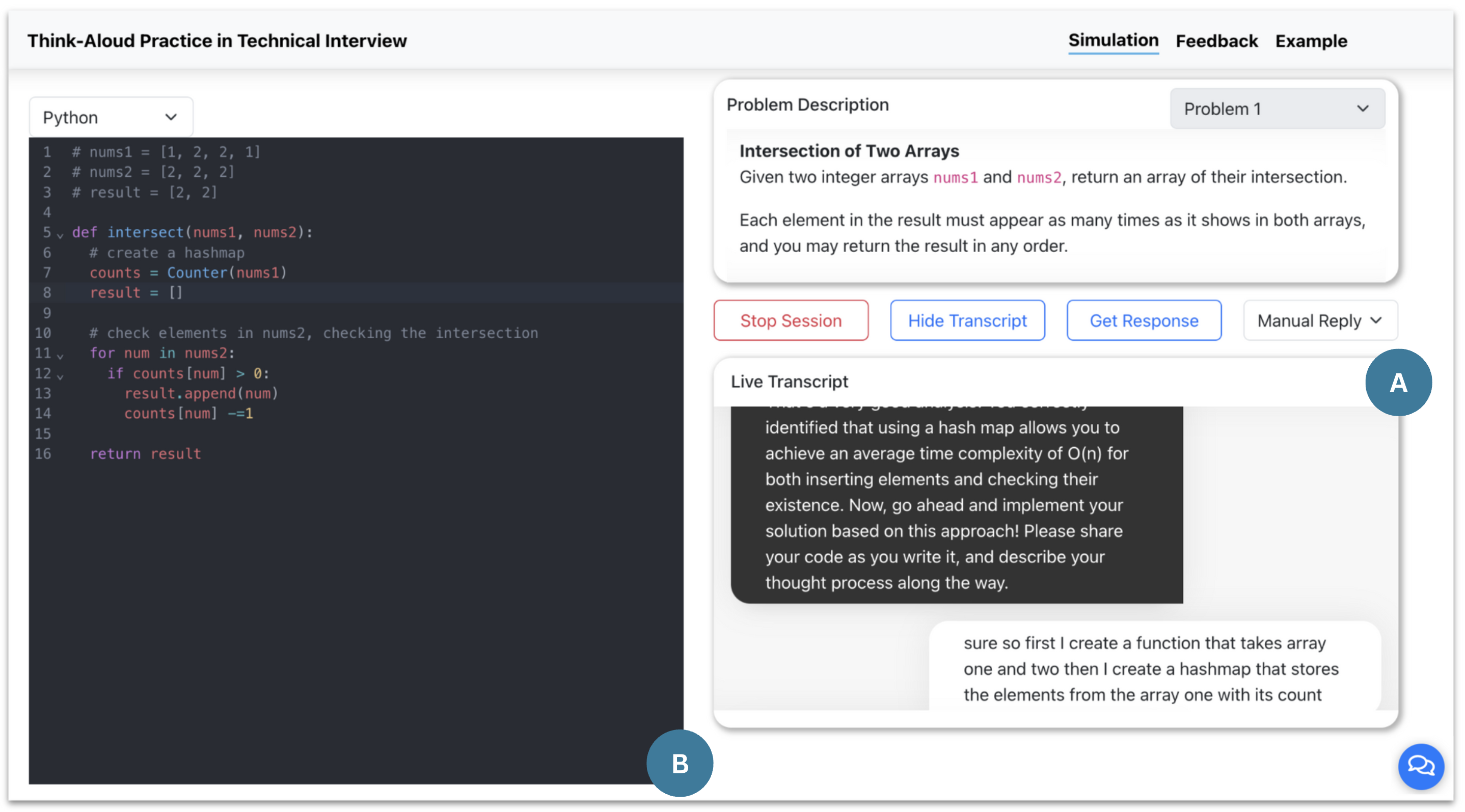}  
    \caption{\textbf{\TSimulation{Technical Interview Simulation}}. AI-facilitated mock interview tool that simulates a technical interview experience through (A) voice-based natural conversation. The interaction flow mimics an actual technical interview, beginning with the AI interviewer asking questions, followed by user responses. During this interaction, users can type code in an (B) integrated editor while verbally \textit{think-aloud} explaining their thoughts and solutions.}
    \label{fig:simulation_feature}
\end{figure*}

\subsection{Facilitating Experiential Learning Through Conversational AI}

Kolb's Experiential Learning theory provides a framework for understanding and structuring skill development \cite{kolb2014experiential}, making it highly relevant to the design of AI systems for think-aloud practice in technical interviews. By emphasizing \textit{learning by doing}, the theory aligns closely with the iterative nature of practicing technical interviews, where candidates gain expertise through cycles of engagement, reflection, and improvement \cite{kolb1974experiential}. The theory’s four-step Experiential Learning cycle --- Concrete Experience, Reflective Observation, Abstract Conceptualization, and Active Experimentation --- offers a structured approach to developing both technical and communication skills \cite{kolb2014experiential, chan2012exploring}, which are critical in the think-aloud process. 


The theory's learning cycle begins with \textit{Concrete Experience}, where learners engage directly in a task, confronting challenges and building hands-on familiarity with the skills they aim to develop \cite{lewis1994experiential}.
For learning to be meaningful, the experience should be \textit{concrete}, which means that learners should immerse themselves in contextually relevant experiences \cite{morris2020experiential}. After participating in a concrete experience, learners move to the \textit{Reflective Observation} phase, where they are encouraged to critically assess their performance \cite{kolb2014experiential}. Following this, they then move on to the next phase, which is \textit{Abstract Conceptualization}. In this step, learners engage in a sense-making process where they connect their observations with broader knowledge to form a conceptual understanding of what they should improve \cite{morris2020experiential}. Finally, the last phase, \textit{Active Experimentation}, allows learners to re-engage with the experience by applying what they have learned and testing their abstract conceptualization \cite{morris2020experiential}. Overall, this iterative learning process enables users to improve their skills through repeated practice.

This four-stage experiential learning process reflects the core principles highlighted by recent studies, such as supporting learning in computing concepts \cite{obrenovic2012rethinking}, STEM education applications \cite{hsu2021possible}, and developing new technical skills \cite{talone2017enhancing, girouard2021reducing}. Furthermore, recent studies have demonstrated the potential of conversational AI in supporting experiential learning \cite{lan2024developing, lin2024enhancing}. Building on these perspectives, our study designs a conversational AI-based technology to facilitate think-aloud practice, ensuring that each step of Kolb’s experiential learning cycle is integrated into technical interview preparation. 


\section{System Design}

\begin{figure*}[t] 
    \centering
    \includegraphics[width=.75\textwidth]{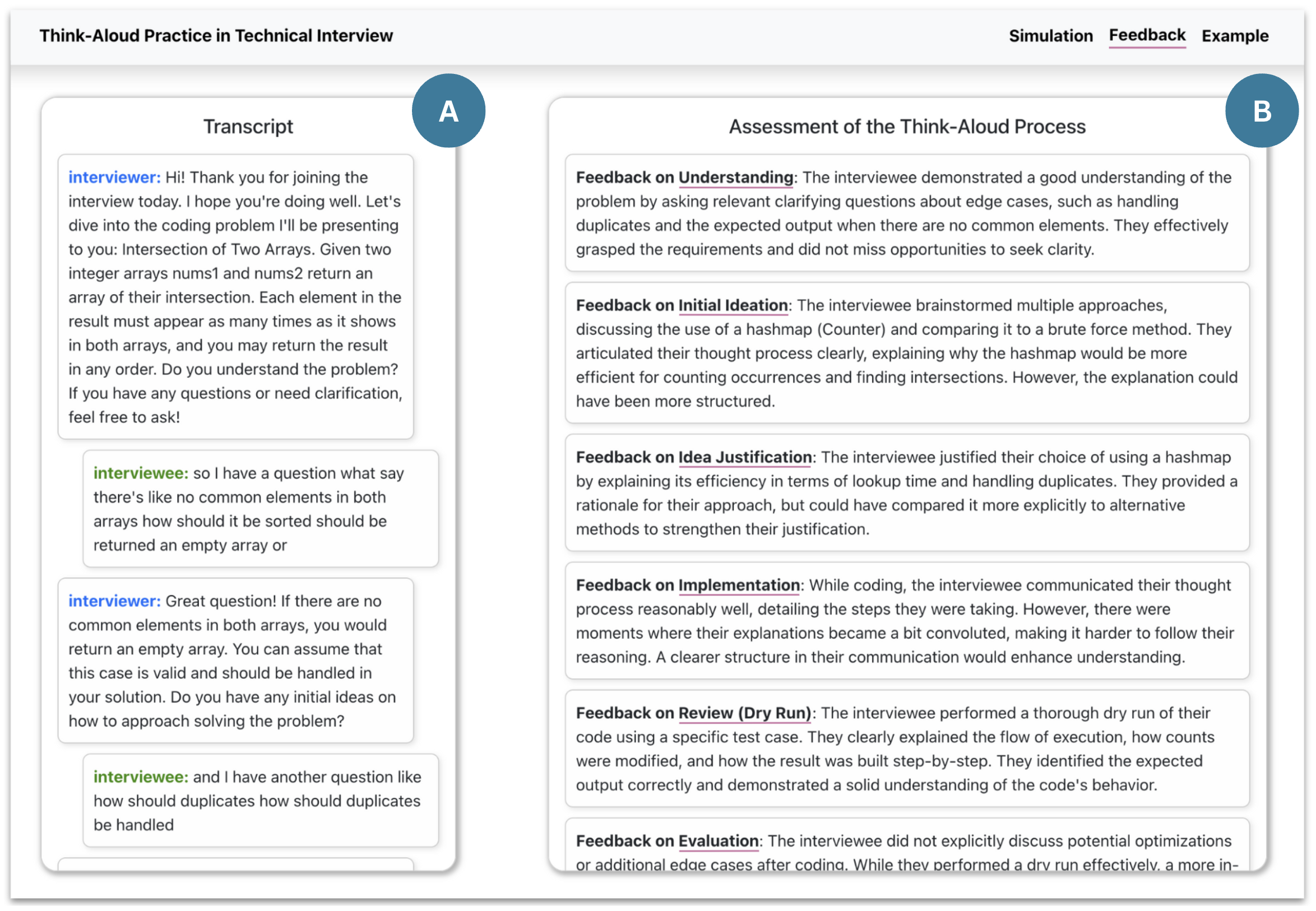}  
    \caption{\textbf{\TFeedback{AI Feedback on Think-Aloud Practice}}. Providing (B) feedback  on users' think-aloud performance based on the interview simulation (A) transcript.}
    \label{fig:feedback_feature}
\end{figure*}

We developed an LLM-based technical interview practice tool to better elicit participants' insights during a user study on how AI can support think-aloud practice in technical interview preparation. Our tool has three main features: \TSimulation{Technical Interview Simulation} (Fig. \ref{fig:simulation_feature}), \TFeedback{AI Feedback on Think-Aloud Practice} (Fig. \ref{fig:feedback_feature}), 
\newline
and \TExample{AI-Generated Think-Aloud Example Dialogue} (Fig. \ref{fig:example_feature}). These features are developed in alignment with Kolb's Experiential Learning theory \cite{kolb2014experiential}. The user flow begins with the \textbf{simulation} feature, where users participate in a mock technical interview with the AI interviewer. Here, users can practice answering coding questions by thinking aloud verbally while coding, replicating an actual interview experience to gain \textit{concrete experience} \cite{kolb2014experiential}. After completing the simulation, users can use the \textbf{feedback} feature to review AI-generated feedback on their think-aloud performance during the interview as part of their \textit{reflective observation} process \cite{kolb2014experiential}. Following this, users can access the \textbf{example} feature to learn from AI-generated example dialogues, which illustrate think-aloud techniques through a sample interaction between interviewer and interviewee, allowing \textit{abstract conceptualization}.  Finally, users can return to the \textit{simulation feature} to practice the same or a new coding problem, applying insights gained from the feedback and examples as their way of \textit{active experimentation} \cite{kolb2014experiential}. Our tool is implemented as a functional web application with initial design decisions for each feature informed by prior literature, as described in the related work and elaborated further in the following subsections.




\subsection{\textbf{Technical Interview Simulation}}

\TSimulation{Technical Interview Simulation} introduces an AI-facilitated mock interview system that simulates a technical interview experience through voice-based natural conversation (Fig. \ref{fig:simulation_feature}). This approach aims to provide users with a \textit{concrete experience} \cite{kolb2014experiential} in technical interviews by replicating the dynamic interaction between a candidate and an interviewer, where the candidate is expected to think aloud during the interview. 

Leveraging OpenAI's GPT-4o-mini, along with \textit{text-to-speech} and \textit{speech-to-text} models, we prompted the AI interviewer to ask a technical interview question and respond to the user’s communication during the think-aloud process. We designed the prompt and interaction flow to mimic an actual technical interview based on prior literature \cite{mcdowell2013cracking, codepath2024umpire}. The interaction begins with the AI interviewer asking a coding question, followed by user responses. During this interaction, users can also type code in an integrated editor while verbally explaining their thoughts and solutions. The AI interviewer is prompted to respond contextually, considering both the conversation and the code typed by the user.

\subsection{\textbf{AI Feedback on Think-Aloud Practice}}
To provide an evaluation of the users' think-aloud performance following the interview simulation, we implemented an \TFeedback{AI Feedback on Think-Aloud Practice} (Fig. \ref{fig:feedback_feature}). This feature facilitates \textit{reflective observation}, a key component of experiential learning \cite{kolb2014experiential}, which helps users reflect on their performance. To generate the feedback, we leveraged LLM by inputting the entire transcript of the participant's technical interview with the AI from the simulation stage.

Our LLM-based tool was designed to provide feedback into six sections, mirroring the typical think-aloud process as established in prior literature \cite{mcdowell2013cracking, ford2017tech} and industry-recommended practices \cite{codepath2024umpire}. Each section represents a key cognitive step in technical interviews:

\begin{enumerate}
    \item \textbf{Understanding}: Evaluate whether the interviewee correctly expressed their understanding of the question by asking relevant clarifying questions and illustrating with a sample test case \cite{mcdowell2013cracking, codepath2024umpire}.
    \item \textbf{Ideation}: Assess how the interviewee brainstormed initial ideas and solutions \cite{mcdowell2013cracking}. It is acceptable if candidates start with a brute-force solution and then optimize their idea gradually \cite{mcdowell2013cracking}.
    \item \textbf{Idea Justification}: Evaluate how the interviewee justifies their approach and walks through their solution before implementing it \cite{mcdowell2013cracking}.
    \item \textbf{Implementation}: Provide feedback on how well the interviewee communicated their thought process and approach clearly while coding \cite{mcdowell2013cracking, codepath2024umpire, ford2017tech}.
    \item \textbf{Review}: Assess how the interviewee reviews their code with a test case by conceptually walking through the code logic \cite{mcdowell2013cracking, codepath2024umpire}.
    \item \textbf{Evaluation}: Provide feedback on how the interviewee evaluated the time and memory complexity of their code and identified potential improvements \cite{mcdowell2013cracking, codepath2024umpire}.
\end{enumerate}



\begin{figure*}[t] 
    \centering
    \includegraphics[width=.85\textwidth]{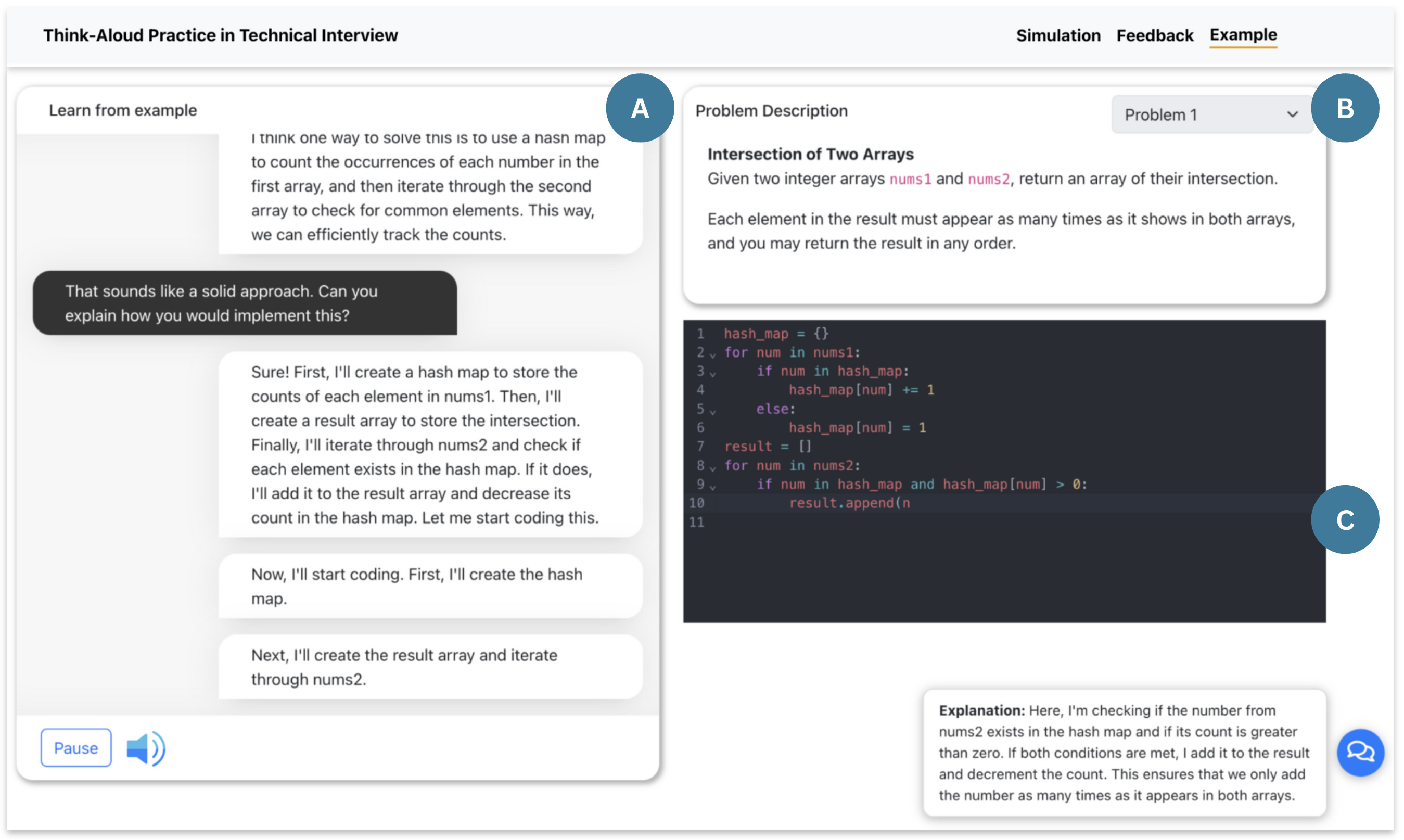}  
    \caption{\textbf{\TExample{AI-Generated Think-Aloud Example Dialogue}}. Providing (A) AI-generated dialogues that model thinking aloud during a technical interview for each (B) coding problem. The goal is to help users learn from examples relevant to their current task and better articulate their thought processes.  The feature also includes a step-by-step (C) code solution alongside the dialogue, with each utterance displayed sequentially and accompanied by voice output.}
    \label{fig:example_feature}
\end{figure*}

\subsection{\textbf{AI-Generated Think-Aloud Example Dialogue}}
\TExample{AI-Generated Think-Aloud Example Dialogue} provides AI-generated dialogues to demonstrate thinking aloud in a technical interview for each coding problem (Fig. \ref{fig:example_feature}). The primary goal of this feature is to help users learn from examples directly related to their current coding practice, allowing them to gain insights into how to articulate their thought processes clearly \cite{ford2017tech}. Additionally, this feature presents a step-by-step code solution alongside the dialogue to show the solution to the user as well \cite{behroozi2019hiring}.

The concept of learning through observing examples aligns with the experiential learning process \cite{hoover2012eyes}. While concrete experience is important for experiential learning \cite{kolb2014experiential}, individuals do not necessarily have to participate in the experience directly \cite{bandura1986social, hoover2012eyes, nadler2003learning}. Research has shown that learning can occur vicariously \cite{roberts2010vicarious}, where individuals learn by watching others. As such, users can gain experience through observation and learn from it. Typically, vicarious learning occurs by observing real people \cite{roberts2010vicarious}. For example, individuals can watch mock interview examples on platforms like YouTube~\cite{bell2025}. However, given the thousands of available coding interview problems \cite{leetcode}, not every problem has an associated think-aloud example available on YouTube. Furthermore, most coding practice platforms like LeetCode \cite{leetcode} and HackerRank \cite{hackerrank} provide only the coding solution without guidance on verbalizing the thought process for a given problem. To address this gap, we propose using AI-generated think-aloud example dialogues as an alternative learning resource. By providing a problem description, we generated example technical interview dialogues between an interviewee and an interviewer using an LLM, as previous research has shown that LLMs can produce contextually relevant conversations \cite{abbasiantaeb2024let}.

We prompted the LLM to generate example dialogues between an interviewee and interviewer in the context of a technical interview, given the input problem, where the interviewee answers the coding problem by thinking aloud. In our prompt, we specified that the interviewee should follow best practices for technical interviews \cite{behroozi2019hiring, behroozi2020debugging, mcdowell2013cracking, codepath2024umpire}, including clearly explaining their approach \cite{behroozi2019hiring}, considering edge cases \cite{mcdowell2013cracking}, asking clarifying questions as needed \cite{mcdowell2013cracking, codepath2024umpire}, and following the typical steps in technical interview:  \textit{understanding}, \textit{initial ideation}, \textit{idea justification}, \textit{implementation}, \textit{review}, and \textit{evaluation} \cite{codepath2024umpire, mcdowell2013cracking}. Additionally, we specified that the interviewer in the dialogue should respond to the interviewee and ask guiding questions to guide the interviewee's thought process \cite{behroozi2020debugging}. 



\subsection{Implementation Notes}
The tool is implemented as a web application using the VueJS framework \cite{vuejs}, along with HTML, CSS, and TypeScript. The backend components are built using the Flask framework \cite{flask} with Python. For all LLM usage in the tool, we use OpenAI's GPT-4o-mini \cite{gpt4omini} due to its low latency \cite{gpt4omini} and good performance \cite{chatbot_arena}, making it suitable for conversational AI implementation in our tool. For the speech-to-text model, we use the Web Speech API \cite{web_speech_api} to convert users' voice input into text, which is then fed into the LLM to generate responses. We use OpenAI's ``tts-1" text-to-speech model \cite{openaitts} to convert the text responses into audio output. All prompts used for the LLM are provided in \textbf{Appendix A}.


\section{User Study}
\subsection{Participants} 
We recruited participants by distributing a screening questionnaire to collect information about their prior experience with technical interviews and their demographic background. Eligible participants were required to have some experience with technical interview practice and be at least 18 years old.

A total of 17 CS students from the first author’s university participated in the study (10 men, 7 women; ages 19–32, median age 21). Thirteen were undergraduate students (7 in their third year and 6 in their fourth year), and four were doctoral students. Participants’ experience with technical interviews varied: 3 had no prior experience, 10 had completed between one and five interviews, 3 had completed six to ten, and 1 had completed more than ten. In terms of race and ethnicity, participants identified as Asian (8), White (6), Middle Eastern/North African (2), and Hispanic/Latino (1). A detailed breakdown of participant demographics is provided in \textbf{Appendix B}.


\subsection{Procedure}
Our research procedure was approved by the Institutional Review Board (IRB) at the first author’s institution. We first began by explaining and demonstrating our tool \textit{(10 minutes)}. After this, we began the actual study by asking participants to engage in the following steps when using the tool:

\begin{enumerate}
    \item First, participants used the \textit{simulation feature} \textit{(15 minutes)} to practice answering a coding question as if in an actual interview. They engaged in a back-and-forth conversation with the AI interviewer while simultaneously typing their code. More detailed information regarding the coding problems are provided in \textbf{Appendix C}.
    \item Next, they used the \textit{feedback feature} \textit{(10 minutes)} to review feedback on their think-aloud performance.
    \item Then, participants accessed the \textit{example feature} \textit{(10 minutes)} to learn from AI-generated think-aloud example dialogues, which demonstrated think-aloud processes for the problem they previously solved in the simulation.
    \item Finally, participants returned to the \textit{simulation feature} again \textit{(15 minutes)} to answer another coding question so that they had a chance to apply what they had learned.
\end{enumerate}

After using the tool, we conducted semi-structured interviews with each participant. We asked participants questions to gather insights on their perspectives regarding using AI for think-aloud practice in technical interviews, as well as any design considerations, concerns, and challenges they encountered. At the end of the user study, participants were compensated with a \$20 Amazon e-gift card.

\subsection{Data Collection and Analysis}
All user studies were conducted via Zoom, with each session recorded and transcribed. We used thematic analysis \cite{braun2021thematic, braun2023toward} to analyze the interviews. The first author conducted an initial round of inductive coding to construct codes grounded in the data. These codes were then reviewed and discussed with another author to refine the coding. The first author then applied the refined codes to all transcripts. Finally, we constructed broader themes and connected them to the tool's features.



\section{Findings}

    


\subsection{Technical Interview Simulations}
\subsubsection{Facilitating Think-Aloud Practice Through Conversational Turn-Taking} \label{5_1_1}

During \TSimulation{Technical Interview Simulation}, participants felt that engaging in turn-taking with the AI created a sense of social presence, allowing them to naturally verbalize their thoughts as part of a two-way conversation. Unlike their prior methods of explaining their thought process to themselves in isolation, they found that engaging in a dialogue with the AI made the experience more interactive and realistic, similar to an actual technical interview (P1, P3, P5, P6, P9, P11, P12). As P1 explained from his experience:

\begin{quote} \textit{"Usually the way I [practice] think-aloud is I would just pretend that there's someone in my mind and I will pretend that I'm trying to convey something, but I don't know whether it properly understands me. [However, conversing with AI] is more like a two-way communication. So [this] interaction makes it more engaging and doesn't feel like I'm just trying to shadow interview"} - P1 \end{quote} 

As such, participants felt the interactive responses from AI during the simulation felt like quick feedback, which they perceived as helping them understand whether their explanations could be properly understood. Moreover, participants felt that the interactivity of the AI made the interaction feel more realistic and less boring. As P2 mentioned, "The realism of the simulation really felt like I was in an interview [and] I wasn't bored.
" Additionally, such interaction also helped participants keep the think-aloud process on track (P13).


Furthermore, participants also highlighted the need for additional features that help them convey their explanations more clearly. They suggested incorporating additional modalities, such as allowing users to highlight sections of their code and provide verbal explanations linked to those specific highlights (P7, P8). P8 noted that, in a real interview, "most programmers would probably [want] to highlight their code and explain the process [to the interviewer] just to make it more specific."

\subsubsection{Concerns About Overly Positive AI Interviewers}\label{5_1_2}

While participants generally appreciated the simulation, some expressed concerns that the AI interviewer tended to always respond positively regardless of the quality of their explanation (P1, P3–P6, P12), noting that interviewees might encounter a harsher interviewer in an actual interview. For example, P1, who had completed more than 10 technical interviews, reflected:
\begin{quote}
    \textit{"You can get harsher interviewers where they will not give you enough things. So in those cases, that's something you would also want to practice"} - P1
\end{quote}

P1 further noted that practicing with a friendly AI interviewer might create a gap in expectations between the simulation and an actual interview where "[interviewee] might expect certain things that an AI model is giving by being humble ... [but] in a real interview, that could be challenging" (P1). To address this, several participants (P1, P3–P6, P12) suggested adding an option for customizable AI personas to mimic diverse interviewer styles, including stricter or less engaging behaviors.


\subsection{AI Feedback on Think Aloud Practice}

\subsubsection{Providing Feedback Beyond Verbal Content Analysis} \label{5_2_1}

Through interactions with our tool, participants identified several aspects that \TFeedback{AI feedback on think-aloud practice} should address. For instance, participants found that breaking down the feedback for each think-aloud step helped them pinpoint specific areas for improvement (P2, P5, P7, P9, P12, P16). In addition to descriptive feedback, participants also suggested including feedback on the use of filler words (P1, P16), providing score-based feedback (P4, P6), and addressing pauses during the think-aloud process (P10). For example, P10 pointed out that he often got stuck and remained silent for several seconds, suggesting that the AI feedback should account for these pauses. 



Additionally, providing feedback beyond verbal content analysis was also considered important. For instance, P17 mentioned that the tool should provide feedback on \textit{balancing time between thinking, talking, and coding}, as he found this to be a challenge during the think-aloud process:

\begin{quote}
    \textit{"I was having a little bit of difficulty figuring out when I should be thinking and actually coding versus actually like explaining. Cause I wasn't sure like. 'Oh, in this portion should I be coding it, or in this portion should I be like explaining'" - P17}
\end{quote}

To address this, P17 suggested incorporating a timeline that offers recommended time allocations, such as how many seconds users might spend thinking before they start talking and how long they should ideally spend talking at each step.

\subsubsection{Framing Feedback to Improve Trust} \label{5_2_2}
Participants emphasized that the manner in which feedback is presented also plays a critical role. For example, P14 suggested aligning the feedback with what actual interviewers might expect to improve its credibility. Additionally, the way feedback is framed can influence how trustworthy it appears. P14 noted that presenting feedback from a third-person interviewer perspective, rather than directly from the AI, could make it more convincing. For example, P14 suggested that the AI feedback could be phrased as "Interviewers would find this aspect of the interviewee impressive... but there are other qualities interviewers might look for" instead of directly stating that the interviewee is lacking in certain areas.


In terms of feedback explainability, P8 recommended linking the feedback directly to specific user utterances so users could clearly understand which parts of their speech the feedback referred to. He also argued that this approach could help users navigate the specific part of their transcript where they should apply the feedback. 



\subsection{AI-Generated Example of Technical Interview Dialogues}

\subsubsection{Vicarious Learning Through AI-Generated Examples Allow Self-Evaluation} \label{5_3_1}

Participants generally found that \TExample{AI-generated think-aloud example dialogue} facilitated vicarious learning by showing them how to think aloud when approaching technical interview problems and enabling them to compare it to their own approach (P1 – P6, P9, P11, P12). They described several benefits: understanding the thought process (P12), learning effective ways to articulate thoughts (P5), identifying helpful questions to clarify the problem (P2), and understanding the expected structure of interview responses (P9). Additionally, participants mentioned that these examples enabled them to compare it with their own performance, as P2 noted:

\begin{quote} \textit{“It allows me to compare how I did in comparison to how the really good interview should go. So, [previously] seeing how I did, and then [comparing it to] how the AI did it, it's like. 'Oh, I can do that too ... when I do this next problem.' This feature is like a benchmark” - P2} \end{quote}

When compared to simply seeing a coding solution, P11 and P12 viewed AI-generated think-aloud example dialogue between AI interviewer and AI interviewee more favorably than the static solution offered by platforms like LeetCode \cite{leetcode}, which only provides code solutions and logical explanations. P11 also explained that conversation examples are more helpful than “just someone telling the logic,” as they allow users to observe how to think aloud step-by-step. Additionally, P6 mentioned that learning from AI-generated example dialogues felt similar to his experience when observing others perform mock technical interviews, which he found insightful. For beginners learning the think-aloud technique, such examples can also provide "inspiration on how to start and what to do" (P1).

Moreover, P1 emphasized that having AI-generated example dialogue tailored to each specific problem was more beneficial than relying on generic examples or examples from different problems, such as those found on platforms like YouTube. P1 noted that YouTube often lacks specific think-aloud examples for every coding question, and when using examples from different questions, there is an added challenge of transferring that knowledge to the current problem. As P1 explained: "If I get an example on a different interview question, I'll still have to transfer that knowledge and try to apply it to this [question] ... So, having examples on the exact question is really helpful." 

\subsubsection{Unrealistic AI-Generated Examples Can Discourage Learners} \label{5_3_2}

Several participants raised concerns that the AI-generated think-aloud examples were often perceived as too perfect for them (P2, P3, P7, P8), particularly in communication, as they presented an overly organized line of thought (P3). Specifically, P3, a non-native English speaker, mentioned:

\begin{quote} 
\textit{“The answer is too good to be true because I don't think in any real interview, even if you are very good at coding interviews [...] you are going to have this exact, very organized line of thought.” - P3} 
\end{quote}

Therefore, participants perceived that providing realistic think-aloud examples is crucial because a perfect example "makes it hard to adapt." (P3) Similarly, P7, a native English speaker, noted that while the perfection of the example makes it trustworthy, “it feels less like something I would follow.”
Moreover, for P8, seeing a perfect example caused self-doubt, as he felt he couldn’t perform as perfectly as the AI-generated response.

Nevertheless, some participants saw value in striving for the level of perfection provided by AI-generated examples (P1, P6, P11, P16). P1 and P6 emphasized that these examples are fine as long as users are "not necessarily trying to replicate [the] exact... words" (P1). P1 further commented that doing a simulation after seeing the example can be beneficial, as users can try to "explain things differently" based on their communication style.



To mitigate concerns about overly perfect examples, several participants suggested that the examples could be adjusted based on user preferences (P7). Similarly, other participants (P8, P14, P16) suggested that showing multiple alternative responses and providing explanations for why a particular response might be preferred over others could help users better adapt the examples to their own styles. Furthermore, P14 suggested a \textbf{human-AI collaborative crowdsourcing approach }, which basically if the tool were deployed publicly, users could "crowdsource [examples] from their simulations with AI," particularly when they receive positive feedback. He felt this human-AI collaboration approach is better since "the examples come from humans, making them more realistic, and since those examples are also [accompanied by] the AI feedback, people can see if they are good examples."

\subsection{Supporting Inclusion and Addressing Intersectional Challenges in AI-Assisted Interview Practice} 

\subsubsection{Promoting Inclusion and Equal Access Through AI-Assisted Technical Interview Practice} \label{5_4_1}
Participants highlighted that AI-based tools for practicing think-aloud during technical interview preparation can promote equal access to practice opportunities. For instance, P11, an Asian woman, expressed that she does not have many friends of the same gender in computer science with whom she can conduct mock interviews, as she explained:
\begin{quote}
    \textit{“I don't have a lot of female friends who are in computer science or just friends in general and computer science. And so having a tool like this that's free... would allow for more people to have access to [practice].” - P11}
\end{quote}
P11 also mentioned her lack of confidence in technical interviews and her hesitation to appear "vulnerable to [her] friends," preferring instead to practice with AI. She further noted, "Not everyone has that support system. So this tool definitely evens the playing field." 

Additionally, several participants (P9, P11, P12, P13) suggested that integrating AI-assisted technical interview practice into the class curriculum could be beneficial. Despite the importance of practicing think-aloud or technical interviews in general, P11 mentioned, "None of our classes... teach us coding interview things." Participants noted that students usually have to practice independently outside of class, and P9 suggested that incorporating this type of practice into coursework could reduce the “overhead of time commitment” required for technical interview preparation outside the classroom.

\subsubsection{Intersectional Challenges in AI-Assisted Think-Aloud Practice} \label{5_4_2}
During our study, participant P3, a Middle Eastern/North African woman and non-native English speaker, frequently said "sorry" when she mispronounced words or tried to remember coding syntax during the interview simulation. P3 suggested that her tendency to apologize caused the AI to misinterpret her communication as uncertainty about her answers, resulting in feedback indicating she was unsure of her solutions. While the AI’s ability to identify such behavior can be seen as a feature, she expressed that such behavior should be considered acceptable, emphasizing that there should be "room for error" when the AI assesses participants' think-aloud performance. As she noted:

\begin{quote} 
    \textit{“I'm not a native speaker, ... English is not my first language. So sometimes I'm like, ‘Yeah, sorry, blah blah.’ [So] in the feedback, it says, ‘You are very unsure.’ I think it does that because I said, ‘Oh, sorry.’ ‘Oh, no,’ and all of this stuff. But I think [there should be a] room for error.” - P3}
\end{quote}

Similarly, participant P11, an Asian American woman and native English speaker, also tended to apologize during the simulation, such as when taking time to think about how to solve the coding problem. However, unlike P3, P11 appreciated the AI feedback that identified her uncertainty. As she mentioned: {"It did give feedback on the fact that I was hesitating and not confident ... which was really cool to see as feedback."} She also noted that, in an actual interview, she tends to apologize less frequently, attributing the higher occurrence of "sorry" during practice sessions to the more casual nature of the environment.

\section{Discussion}

\subsection{Design Implications for AI-Assisted Think-Aloud Practice in Technical Interview Preparation}


\subsubsection{Designing for Social Presence: Turn-Taking and Emotional Reciprocity in Think-Aloud Practice}

In Human-Computer Interaction (HCI) literature, social presence is often discussed in the context of human-virtual agent interactions \cite{pereira2014improving, leite2009time, biocca2003toward, zhang2024explaining}. Kreijns et al. \cite{kreijns2022social} define social presence as the perception of the realness of the other person in technology-mediated communication. In our study, the \textit{other person} refers to the AI technical interview simulator \cite{pereira2014improving, leite2009time}. In the context of \TSimulation{Technical Interview Simulation}, establishing social presence from the AI mock interviewer is important to facilitate a more natural and engaging interaction, thereby facilitating users' think-aloud practice (\S \ref{5_1_1}). Our findings highlight two key design factors that can potentially improve the AI's social presence in technical interview simulation: \textbf{turn-taking} \cite{biocca2001criteria, biocca2003toward} and \textbf{mutual understanding} \cite{biocca2003toward, biocca2001networked}.

Prior work establishes that \textbf{turn-taking}, the structured coordination of conversational exchanges, forms the foundation of effective human-machine dialogue \cite{skantze2021turn}. In our study, participants reported that turn-taking with the AI interviewer not only improved realism but also facilitated deeper engagement (\S \ref{5_1_1}). This mirrors earlier findings \cite{pereira2014improving, lombard1997heart} on how structured exchanges can improve social presence and potentially increase learner satisfaction \cite{gunawardena1997social}. This back-and-forth interaction enabled users to articulate their thought processes (\S \ref{5_1_1}), providing a \textit{concrete experience} \cite{kolb2014experiential} similar to real-world interviews. Moreover, the immediacy of the simulator’s responses helped participants continuously assess whether their explanations were clear and on track (\S \ref{5_1_1}), thereby reinforcing learning through experience  \cite{slovak2017reflective} throughout the interview process.

Furthermore, \textbf{mutual understanding} occurs when both parties in a conversation understand each other’s intentions \cite{biocca2001networked}. In the context of technical interview simulations, this means not only that the user conveys their reasoning clearly but also that the AI provides responses that accurately reflect the quality of the user's explanations so that the user can interpret the true intent of the AI's responses. However, LLMs often exhibit \textit{sycophantic behavior}, generating responses that prioritize user approval over objective accuracy \cite{perez2022discovering, huang2024trustllm}. This manifested when participants noted the AI interviewer being overly positive (\S \ref{5_1_2}). Although supportive responses can increase social presence \cite{Verhagen2014Virtual, pereira2014improving, zhu2022effects}, consistently overstated positivity can undermine the authenticity that is crucial to genuine interaction \cite{lombard1997heart}. 


This challenge is especially significant in think-aloud practice, where users benefit from accurate assessments of their explanations (\S \ref{5_1_1}). If the AI consistently offers praise, users may struggle to identify weaknesses in their reasoning until they receive post-simulation feedback. Although participants suggested adjustable interviewer personas to address this issue (\S \ref{5_1_2}), such persona-based approaches risk introducing \textit{persona-based sycophancy} \cite{huang2024trustllm}, where the AI’s response is more reflective of a chosen persona than of genuine evaluation. To mitigate LLM sycophancy, researchers have proposed fine-tuning models on carefully curated datasets \cite{wei2023simple}. For technical interview practice, these datasets could combine real interview transcripts with human-annotated examples that provide diverse and balanced conversational examples.

\subsubsection{\TFeedback{AI Feedback on Think-Aloud Practice}}



\paragraph{\textbf{Balancing Time Between Thinking, Talking, and Coding}}
Our study found that some participants struggled to balance their time between thinking, talking, and coding (\S \ref{5_2_1}). In technical interviews, candidates are generally expected not to remain silent for extended periods, as verbalizing their thought process is a critical part of the evaluation \cite{ford2017tech}. When candidates do not immediately find a solution, interviewers typically expect them to articulate their reasoning, which also allows interviewers to offer hints based on the think-aloud process \cite{behroozi2020debugging}. As such, providing feedback when candidates are silent for too long could be valuable. Participants also suggested offering guidance on how to allocate time across different activities during problem-solving (\S \ref{5_2_1}). Enabling this type of feedback may require collecting transcribed technical interview data and training a model to identify ideal time allocations for each stage of specific coding problems \cite{hsu2024piece}.


\textbf{Framing of Feedback: AI as a Mediator, Not Authority.}
Our study revealed novel insights about feedback presentation and trust. Building on research about trustworthy LLM outputs \cite{huang2024trustllm, liu2023trustworthy}, we found that participants perceived feedback as more credible when presented from a third-person interviewer perspective rather than directly from the AI (\S \ref{5_2_2}). For example, framing feedback as an explanation of how some interviewers typically expect certain responses, rather than directly instructing the candidate from a first-person perspective, can potentially increase perceived trustworthiness (\S \ref{5_2_2}). This aligns with prior research that shows presenting information from third-person perspectives can make AI systems appear more objective than authoritative \cite{lee2025countering}. Therefore, in the context of technical interview practice, grounding and framing the AI's feedback as delivering insight from actual real-life hiring managers could be more conducive to the users' learning experience.

\subsubsection{\TExample{AI-Generated Think-Aloud Example Dialogue}}

\textbf{\textit{Crowdsourcing Think-Aloud Examples through Human-AI Collaboration.}}
In order to collect more realistic and diverse examples, our users suggested a crowdsourcing approach (\S \ref{5_3_2}). Crowdsourcing refers to the practice of multiple individuals collectively contributing to achieving a certain goal, such as generating content for learning \cite{Pirttinen2023Lessons, Suhonjic2019Fostering}. To do this, a human-AI collaboration approach can be useful and potentially more efficient \cite{Abad2017Autonomous, choi2024vivid}. For example, when users engage in a technical interview simulation with an AI interviewer, they can opt to share their recorded simulation with other users, especially when they receive positive feedback from the AI. Learning from examples involving real users can make them feel more relatable while still upholding best practices, as each example is accompanied by AI feedback. This collaborative approach also mitigates the risk of hallucinations that can occur when examples are entirely generated by LLM without human involvement \cite{huang2023survey}.

\subsection{Rethinking the Role of AI in Technical Interview Practice: Leveraging Human-AI Collaboration}

To broaden our discussion, we situate our findings within the broader field of AI-assisted interview practice. Previous work in this domain has primarily focused on developing AI interviewers as stand-alone practice partners, designed to simulate human-like interactions during mock interviews \cite{daryanto2024conversate, TARDIS, inoue2020jobErica2, rasipuram2020automatic, naim2016automated, Chou2022-mm, thakkar2023automatic, li2023ezinterviewer}. While such systems can provide accessible practice opportunities for those lacking human partners \cite{inoue2020jobErica2, rao2021FollowUpQuestionGenerator, li2023ezinterviewer, daryanto2024conversate}, this raises an important question: \textbf{should the primary goal of these systems be to entirely replace human practice partners?}

While such an AI could benefit those who struggle to find practice partners \cite{daryanto2024conversate, hoque2013mach} (a key premise of our study), alternative roles for AI \cite{wilson2018collaborative} might also be helpful for interview practice. This is particularly relevant for individuals who prefer mock interviews with other humans. For instance, P17 mentioned that while he valued practicing with an AI interviewer, he preferred practicing with friends because he found it more enjoyable and experienced a stronger human connection. However, he noted the drawbacks of practicing with friends, which can be “more laid back because there is no one pressuring you [and that] they usually don’t give great feedback.” Given this example, how might we best leverage AI to enhance interview practice without entirely replacing the role of human partners?

\textbf{One potential direction could be leveraging human-AI collaboration} that leverages the potential of both humans and AI \cite{kim2022StudentAIcollaboration, zhang2020effectAIAsisstedDecisionMaking}. For example, as P17 mentioned:
\begin{quote}
  \textit{  "Instead of just relying on the AI ... you can basically just have the interview going on between the two people. And then, the AI interviewer goes over their feedback ... because [the AI feedback] already has a lot of detailed information that extends further into the actual code as well as the talk aloud portion. So I think that helps bring both worlds together because you're getting both the human interaction as well as the AI portion" - P17}
\end{quote}



Future research should not only investigate additional applications of this collaborative approach but also evaluate its effectiveness compared to practice sessions conducted exclusively with either AI or human partners.

\subsection{Leveraging AI-Assisted Technical Interview Practice to Promote Inclusivity in Computing Careers}

Technical interviews are challenging for all job applicants in computing \cite{behroozi2019hiring}. These challenges can be even more pronounced for individuals minoritized in the field \cite{hall2018effects, yamaguchi2019intersectionality, lunn2022need, zavaleta2022additional}. This highlights the importance of creating inclusive and accessible resources, such as through AI-based technical interview practice (\S \ref{5_4_1}). While an AI tool provides valuable support, it does not entirely address the unique challenges faced by students who are minoritized in computing in this context. Research shows that emotional support and peer connections remain crucial for career success in computing fields \cite{Weaver2021CareerRelated}. Although AI can provide certain forms of emotional support \cite{Merrill2022AI}, it currently struggles to replicate the nuanced emotional interactions that humans provide \cite{Meng2023Mediated}. Moreover, relying solely on AI for practice can unintentionally promote individualism \cite{okun2000white}, potentially reducing the collaborative support networks that are especially beneficial for minoritized groups \cite{lunn2024you}. These limitations suggest the need for balanced resources that integrate AI with community-based support. 

Additionally, when designing AI-based technical interview tools, it is important to consider the intersectional challenges that some users may face (\S \ref{5_4_2}). Intersectionality refers to the way that individuals' various social identities, such as race, gender, and language background, may overlap to create unique experiences shaped by their privilege and discrimination \cite{crenshaw2013mapping}. These intersecting identities may also influence how users interact with AI-based interview training tools. For example, one participant, a Middle Eastern/North African woman and non-native English speaker, frequently apologized during the interview simulation (\S \ref{5_4_2}). Through the lens of gender identity, research indicates that women are often socialized to apologize more frequently \cite{Holmes1995Women}, especially in professional settings \cite{Hobbs2003The}, which may be linked to societal expectations surrounding politeness \cite{Holmes1995Women} and perfectionism \cite{okun2000white}. This tendency can intersect with language-based insecurities, especially for non-native English speakers who may experience linguistic challenges and lack confidence when speaking \cite{Tantri2023The}. Moreover, studies have highlighted the underrepresentation of Middle Eastern/North African women in computing \cite{khreisat2009under, islam2019science}. Hence, these compounding factors can make the practice experience more challenging for such users. Additionally, users' insecurities may be amplified when the AI provides critical feedback without accounting for these intersectional factors (\S \ref{5_4_2}). This can discourage users from further engagement \cite{Grundmann2020When}, which can hinder the experiential learning process \cite{kolb2014experiential}. 


However, while researchers can design AI practice systems to account for intersectional factors and mitigate biases, marginalized individuals may still encounter biases during real technical interviews \cite{latu2015gender}. Therefore, future research should focus on designing AI systems that not only reduce bias during practice but also equip users with strategies to navigate biases in real technical interviews. Furthermore, it is important to note that these challenges are not solely the result of individual identity-based experiences but are also shaped by larger systemic issues, including racism, sexism, and xenophobia, which persist in hiring practices and workplace cultures \cite{purkiss2006implicit}. Hence, future work should also focus on strategies to promote fairer hiring practices.

\section{Limitations}
By having users interact with our tool, we gained insights into users' preferences and design considerations for using AI in think-aloud practice for technical interview preparation. While this approach helped identify key user preferences and design opportunities, deploying the tool in a longitudinal study, where users engage with it over an extended period, could provide a deeper understanding of its efficacy and reveal additional design insights. Additionally, this study included a limited subset of participants from a single university in the United States, all of whom were undergraduate or graduate computer science students. Although participants varied in their technical interview experience and demographic backgrounds, the findings may not be fully representative of students from all institutions or backgrounds. Furthermore, while technical interviews often follow a similar format, there may be variations across companies and roles that were not captured in this study.

\section{Conclusion}

Our study explored users’ perceptions of using conversational AI for think-aloud practice, focusing on simulations, feedback, and learning from AI-generated examples. We identified several design recommendations, for example, promoting social presence for technical interview simulation, providing feedback beyond verbal content analysis, and crowdsourced think-aloud examples enabled by human-AI collaboration. Additionally, we explored some intersectional challenges and discussed how AI-assisted technical interview preparation could promote inclusivity in computing careers. Lastly, by connecting our findings to the broader literature on interview practice, we discussed the need to rethink the role of AI in technical interview practice, suggesting a research direction that leverages human-AI collaboration.

\bibliographystyle{IEEEtran}
\bibliography{references}

\vspace{12pt}
\appendix

\subsection{Prompts}

\subsubsection{Prompts for Interview Simulation}

\paragraph{{Prompt for the First Interviewer Message}}

You are a hiring manager conducting a coding interview. Your goal is to assess the candidate's problem-solving skills, coding ability, and communication. Begin by greeting the candidate and presenting the following coding problem. 

\textit{Problem:}
\begin{verbatim}
    [input problem]
\end{verbatim}
\paragraph{Prompt for the Rest of the Interviewer Interaction}

You are a hiring manager conducting a coding interview. Your goal is to assess the candidate's problem-solving skills, coding ability, and communication. Ensure that your communication is short and concise. Provide hints only if the candidate is stuck, but avoid giving too many hints, especially those that reveal the solution.

\textit{The coding problem is the following:}
\begin{verbatim}
    [input problem]
\end{verbatim}

The candidate will implicitly follow these six steps:
\begin{enumerate}
    \item {Understanding:} The candidate may ask clarifying questions to ensure they fully understand the problem and may propose an initial test case to demonstrate their understanding of the requirements.
    \item {Initial Ideation:} The candidate will brainstorm initial ideas on how to solve the problem.
    \item {Idea Justification:} The candidate will justify their approach, explaining why the chosen solution is suitable.
    \item {Implementation:} The candidate will code the solution while thinking aloud to describe their thought process.
    \item {Review (Dry Run):} After coding, the candidate will dry-run their code with a test case, walking through the logic step by step.
    \item {Evaluation:} The candidate will evaluate their solution, discussing possible optimizations, edge cases, and any necessary improvements.
\end{enumerate}

Throughout the interview:
\begin{itemize}
    \item Prompt the candidate to think aloud and explain their reasoning at each step.
    \item Ask follow-up questions to gauge their understanding and depth of knowledge.
    \item Provide hints only if the candidate is stuck, but avoid giving too many hints, especially those that reveal the solution.
    \item Ignore minor typos or grammar errors in the candidate’s responses.
    \item Keep your communication short and concise.
\end{itemize}

At any point, you may refer to the candidate’s current code (ignoring syntax errors and focusing on the logic):

\textit{Candidate's Current Code:}
\begin{verbatim}
    [Code]
\end{verbatim}

Make sure your communication is short and concise.

\texttt{[APPEND CONVERSATION]}

\subsubsection{Prompt for AI-Feedback}
I have a transcript of a coding interview where the interviewee is required to think aloud while solving a problem. Based on the transcript provided below, please give detailed feedback on the interviewee's performance, focusing on the following six aspects. Format the response in JSON using the following structure:

\begin{lstlisting}[basicstyle=\small\ttfamily, breaklines=true]
{
    "understanding" : "Feedback on how the interviewee demonstrated their understanding of the problem by asking clarifying questions and/or proposing a test case.",
    "initial_ideation": "Feedback on how the interviewee brainstormed and proposed their initial ideas.",
    "idea_justification": "Feedback on how the interviewee justified their proposed approach and explain the walk through of their idea.",
    "implementation": "Feedback on how the interviewee communicated their thought process while coding the solution.",
    "review_dry_run": "Feedback on how the interviewee performed a dry run of their code, explaining the execution flow using test cases.",
    "evaluation": "Feedback on how the interviewee evaluated their solution, including discussions of potential optimizations, edge cases, or improvements."
}
\end{lstlisting}

Make sure your feedback is constructive and suggests things that they need to improve for each section. Please also ignore typos or transcription errors.

Here are the specific phases to assess:

\begin{enumerate}
    \item \textbf{Understanding:} Evaluate whether the interviewee correctly expressed their understanding of the question by asking relevant clarifying questions and illustrating with a sample test case. Did they fully grasp the requirements? Were there missed opportunities to seek more clarity?

    \item \textbf{Initial Ideation:} Assess how the interviewee brainstormed initial ideas and solutions. It is acceptable if candidates start with a brute-force solution and then optimize their ideas gradually. Did they consider multiple approaches or stick with a single idea? Did they clearly explain their thought process?

    \item \textbf{Idea Justification:} Evaluate how well the interviewee justified their solution and walk through their solution before implementing it. Did they explain why their approach was suitable or compare it to alternative solutions? Were they able to defend their choice of data structures, algorithms, or logic?

    \item \textbf{Implementation:} Provide feedback on how well the interviewee communicated their thought process while coding. Was their reasoning easy to follow? Were there gaps in their explanation or places where communication became unclear?

    \item \textbf{Review (Dry Run):} Analyze how the interviewee performed a dry run of their code with a test case. Did they clearly explain the flow of execution, identify potential issues, or spot logical errors?

    \item \textbf{Evaluation:} Provide feedback on how the interviewee evaluated their solution after coding. Did they discuss possible optimizations or improvements?
\end{enumerate}

If the interviewee did not perform a phase, note that it was not done. Make sure your feedback is constructive and that you ignore typos or transcription errors.

\textit{Transcript:}
\begin{verbatim}
    [transcript]
\end{verbatim}
\subsubsection{Prompt for AI-Generated Example Dialogue}

Simulate a realistic coding interview between an interviewer and an interviewee focused on solving the given problem. The simulation should include explanations of best practices based on the example provided. When writing the code, present it as "lines of code" and provide step-by-step explanations. Avoid using built-in libraries if possible.

Format your response in JSON format:
\begin{lstlisting}[basicstyle=\small\ttfamily, breaklines=true]
[
    {
        "role (interviewer/interviewee)": "",
        "content": "",
        "code (containing STEP-BY-STEP code that the interviewee writes while thinking aloud)": "",
        "explanation (explain the importance or rationale of answering in such a way)": ""
    },
    ...
]
\end{lstlisting}

The interviewer may implicitly ask probing questions to understand the interviewee's thought process, while the interviewee should clearly explain their approach, consider edge cases, and write code to solve the problem. The interviewer may offer hints if the interviewee gets stuck. 

The interviewee should think aloud, ask clarifying questions if needed, and demonstrate their problem-solving skills.

\textit{Problem:} \texttt{\{input problem\}}

The interview should implicitly follow these six steps:
\begin{enumerate}
    \item \textbf{Understanding:}
    \begin{itemize}
        \item The interviewer introduces the problem and provides examples.
        \item The interviewee may ask clarifying questions to ensure a full understanding of the problem.
        \item The interviewee may propose an initial test case to demonstrate their understanding of the requirements.
    \end{itemize}
    \item \textbf{Initial Ideation:}
    \begin{itemize}
        \item The interviewee brainstorms initial ideas on how to solve the problem.
    \end{itemize}
    \item \textbf{Idea Justification:}
    \begin{itemize}
        \item The interviewee justifies their chosen approach, explaining why it is suitable for the problem.
        \item The interviewee explains the walkthrough of the solution
    \end{itemize}
    \item \textbf{Implementation:}
    \begin{itemize}
        \item The interviewee writes the code to solve the problem, explaining their logic as they go.
        \item The interviewer may ask the interviewee to consider different cases, such as empty inputs or edge scenarios.
    \end{itemize}
    \item \textbf{Review (Dry Run):}
    \begin{itemize}
        \item After coding, the interviewee dry-runs their code with provided examples, walking through the logic step by step.
        \item The interviewee considers additional test cases to ensure robustness.
    \end{itemize}
    \item \textbf{Evaluation:}
    \begin{itemize}
        \item The interviewee evaluates their solution, discussing possible optimizations, edge cases, and any necessary improvements.
        \item The interviewer provides feedback on the solution, highlighting strengths and areas for improvement.
        \item The interviewee reflects on their performance and discusses what they could have done differently.
    \end{itemize}
\end{enumerate}

\newpage

\subsection{Participant Demographics}

\begin{table}[h]
\caption{Participant Demographics}
\small
\begin{threeparttable}
\begin{tabular}{c|c|c|p{1.5cm}|p{2cm}|p{1cm}}
\hline
ID & Gender & Age & Occupation & Race and Ethnicity & Interview Count \\
\hline
P1 & Male & 32 & PhD Student & Asian & \textgreater{} 10 \\
P2 & Male & 20 & 4th Year Undergrad & Middle Eastern & 1--5 \\
P3 & Female & 26 & PhD Student & Middle Eastern & 1--5 \\
P4 & Female & 25 & PhD Student & Asian & 1--5 \\
P5 & Female & 21 & 4th Year Undergrad & Asian & 1--5 \\
P6 & Male & 20 & 3rd Year Undergrad & White & 1--5 \\
P7 & Male & 20 & 3rd Year Undergrad & Asian & 1--5 \\
P8 & Male & 20 & 3rd Year Undergrad & Asian & 1--5 \\
P9 & Male & 20 & 3rd Year Undergrad & White & 1--5 \\
P10 & Male & 19 & 3rd Year Undergrad & Asian & 1--5 \\
P11 & Female & 22 & 4th Year Undergrad & Asian & 6--10 \\
P12 & Male & 20 & 3rd Year Undergrad & White & 6--10 \\
P13 & Female & 21 & 4th Year Undergrad & Asian & 6--10 \\
P14 & Male & 24 & PhD Student & White \& Hispanic/Latino & 0 \\
P15 & Female & 22 & 4th Year Undergrad & White & 0 \\
P16 & Female & 20 & 3rd Year Undergrad & White & 0 \\
P17 & Male & 21 & 4th Year Undergrad & White & 1--5 \\
\hline
\end{tabular}
\begin{tablenotes}
\item *Note: "Interview Count" refers to the number of technical interviews the participant has completed before this study.
\end{tablenotes}
\end{threeparttable}
\label{tab:participants}
\end{table}

\subsection{Coding Tasks for Technical Interview Simulation}

Our technical interview practice tool includes two coding questions for users to choose from. These tasks are adapted from common data structure/ algorithm problems on LeetCode \cite{leetcode}, which are typical in technical interviews \cite{mcdowell2013cracking}. We selected problems that have multiple valid solution approaches, which aligns with our focus on practicing the think-aloud technique, where users explain their thought processes while navigating their solutions. Following guidelines from prior works on technical interview practice \cite{behroozi2022asynchronous, behroozi2020does}, we ensured that the selected tasks were solvable within a reasonable time limit for the user study and neither too trivial nor overly complex. These are the coding tasks:

\subsubsection{Problem 1. Intersection of Two Arrays}

Given two integer arrays \texttt{nums1} and \texttt{nums2}, return an array of their intersection. Each element in the result must appear as many times as it shows in both arrays, and you may return the result in any order.

\paragraph{Example 1}
\begin{quote}
\begin{verbatim}
Input: nums1 = [1,2,2,1], nums2 = [2,2]
Output: [2,2]
\end{verbatim}    
\end{quote}

\paragraph{Example 2}
\begin{quote}
\begin{verbatim}
Input: nums1 = [4,9,5], nums2 = [9,4,9,8,4]
Output: [4,9]
\end{verbatim}
\end{quote}


\textit{Sample Solution 1. Using a Hash Map}
\begin{itemize}
    \item Intuition: Use a hash map to count the occurrences of each number in the first array. Then, iterate through the second array to identify common elements. For each element in the second array, check if it exists in the hash map. If it does, add the element to the result array and decrement its occurrence count in the hash map.
    \item Time Complexity: O(n + m)
    \item Memory Complexity: O(n + m)
\end{itemize}

\textit{Sample Solution 2. Sorting and Two Pointer (Optimizing The Space)}
\begin{itemize}
    \item Intuition: To find the common elements between two arrays, start by sorting both arrays. Then, use two pointers, one for each array, beginning at their respective starting positions. These pointers initially point to the smallest elements in each array. Compare the elements at the two pointers: if they are the same, add the element to the result array and move both pointers forward by one. If the elements are different, increment the pointer pointing to the smaller element, as this ensures we progress toward finding potential matches. 

    \item Time Complexity: O(nlogn + mlogm)
    \item Memory Complexity: O(min(n,m)) (There is no need for additional space for Hash Map)
\end{itemize}

\subsubsection{Problem 2. Two Sum}
Given an array of integer nums and an integer target, return indices of the two numbers such that they add up to the target. You cannot use the same element twice. You can return the answer in any order.

\paragraph{Example 1}
\begin{quote}
\begin{verbatim}
Input: nums = [2,7,11,15], target = 9
Output: [0,1]
\end{verbatim}    
\end{quote}


\textit{Sample Solution 1. Two-pass Hash Table}
\begin{itemize}
    \item Intuition: First, precompute a hash table by iterating through the first array, where each element is added as a key and its index as the value. Then, iterate through the second array. For each element, check if the target complement \text{(target - nums[i])} exists in the hash table. If it does, return the index of the current element and the index of its complement.
    \item Time Complexity: O(n)
    \item Memory Complexity: O(n)
\end{itemize}

\textit{Sample Solution 2. Sorting + Binary Search}
\begin{itemize}
    \item Intuition: The algorithm to solve the Two Sum problem using sorting and binary search involves first pairing each number with its original index and sorting the array by values. Then, for each number in the sorted array, it computes the complement needed to reach the target and performs a binary search in the remaining portion of the array to find this complement. If the complement is found, the original indices of the two numbers are returned. 
    \item Time Complexity: O(n log n)
    \item Memory Complexity: O(n)
\end{itemize}

\end{document}

%% file: macro.tex
\definecolor{buse_skyblue}{rgb}{0.337, 0.706, 0.914}

\definecolor{buse_orange}{rgb}{0.902, 0.624, 0.0}

\definecolor{buse_plum}{rgb}{0.7, 0.3, 0.5}

\definecolor{buse_green}{rgb}{0.4, 0.6, 0.4}

\definecolor{buse_mustard}{rgb}{0.8, 0.7, 0.2} 
\definecolor{buse_red}{rgb}{0.8392, 0.3921, 0.2666}
\definecolor{buse_purple}{rgb}{0.8, 0.475, 0.655}

\definecolor{buse_seafoam}{rgb}{0.5, 0.75, 0.6}
\definecolor{buse_blue}{rgb}{0.3333, 0.3764, 0.6627}

\newtcbox{\TS}{on line,
  arc=3pt,outer arc=3pt,
  colback=buse_skyblue!25!white, colframe=blue!50!black,
  boxsep=0pt,left=2pt,right=2pt,top=2pt,bottom=2pt,
  boxrule=0pt}

\newtcbox{\CS}{on line,
  arc=3pt, outer arc=3pt,
  colback=buse_purple!25!white, colframe=pink!50!black,
  boxsep=0pt, left=2pt, right=2pt, top=2pt, bottom=2pt,
  boxrule=0pt  
}

\newtcbox{\Tone}{on line,
  arc=3pt,outer arc=3pt,
  colback=buse_green!25!white, colframe=green!50!black,
  boxsep=0pt,left=2pt,right=2pt,top=2pt,bottom=2pt,
  boxrule=0pt}

\newtcbox{\Mod}{on line,
  arc=3pt,outer arc=3pt,
  colback=buse_orange!25!white, colframe=orange!50!black,
  boxsep=0pt,left=2pt,right=2pt,top=2pt,bottom=2pt,
  boxrule=0pt}

\newtcbox{\Concern}{on line,
  arc=3pt,outer arc=3pt,
  colback=buse_plum!25!white, colframe=purple!50!black,
  boxsep=0pt,left=2pt,right=2pt,top=2pt,bottom=2pt,
  boxrule=0pt}

\newtcbox{\TSimulation}{on line,
  arc=3pt,outer arc=3pt,
  colback=buse_skyblue!25!white, colframe=blue!50!black,
  boxsep=.5pt,left=2pt,right=2pt,top=2pt,bottom=2pt,
  boxrule=0pt}

  \newtcbox{\TFeedback}{on line,
  arc=3pt, outer arc=3pt,
  colback=buse_purple!25!white, colframe=pink!50!black,
  boxsep=.5pt, left=2pt, right=2pt, top=2pt, bottom=2pt,
  boxrule=0pt}

\newtcbox{\TExample}{on line,
  arc=3pt,outer arc=3pt,
  colback=buse_orange!25!white, colframe=orange!50!black,
  boxsep=0pt,left=2pt,right=2pt,top=2pt,bottom=2pt,
  boxrule=0pt}